\documentclass[a4paper,11pt]{article}
\usepackage{jheppub} % for details on the use of the package, please
\usepackage[T1]{fontenc} % if needed

\title{\boldmath Current-current deformations, conformal integrals and correlation functions}

\author[a]{Gaston Giribet}
\author[a]{Matias Leoni}
\affiliation[a]{Physics Department, University of Buenos Aires FCEN-UBA and IFIBA-CONICET\\Ciudad Universitaria, pabell\'on 1, 1428, Buenos Aires, Argentina.} %\affiliation[c]{A School for Advanced Studies,\\some-location, Country}

\emailAdd{gaston@df.uba.ar}
\emailAdd{leoni@df.uba.ar}
\abstract{Motivated by the recent work on $T\bar{T}$-type deformations of 2D CFTs, a especial class of single-trace deformations of AdS$_3$/CFT$_2$ correspondence has been investigated. From the worldsheet perspective, this corresponds to a marginal deformation of the $\sigma $-model on AdS$_3$ that yields a string background that interpolates between AdS$_3$ and a flat linear dilaton solution. Here, with the intention of studying this worldsheet CFT further, we consider it in the presence of a boundary. In a previous paper, we computed different correlation functions of this theory on the disk, including the bulk 1-point function, the boundary-boundary 2-point function, and the bulk-boundary 2-point function. This led us to compute the anomalous dimension of both bulk and boundary vertex operators, which first required a proper regularization of the ultraviolet divergences of the conformal integrals. Here, we extend the analysis by computing the bulk-bulk 2-point function on the disk and other observables on the sphere. We prove that the renormalization of the vertex operators proposed in our previous works is consistent with the form of the sphere $N$-point functions.}

\begin{document} 
\maketitle
\flushbottom

\section{Introduction}

String theory on AdS$_3$ with NS-NS fluxes admits an exact worldsheet formulation in terms of the $SL(2,\mathbb{R})$ WZW model \cite{MO1, MO2, MO3}. Recently, motivated by new results in the study of integrable irrelevant deformations of CFT$_2$ \cite{Zamolodchikov, Cavaglia} and their possible holographic interpretation, the authors of \cite{Giveon:2017nie} considered a especial class of marginal deformations of the AdS$_3$ theory that induces an irrelevant deformations in the dual CFT$_2$. In terms of the $SL(2,\mathbb{R})$ WZW description, such deformation corresponds to adding to the action an operator quadratic in the Kac-Moody currents; namely 
\begin{equation}
\int d^2z \, J^+\bar{J}^+ \, . \label{Prima}
\end{equation}
This operator, while being marginal in the worldsheet CFT$_2$, induces an irrelevant deformation in the dual CFT$_2$ that can be thought of as a single-trace version of the $T\bar{T}$ deformation \cite{Giveon:2017nie, Giveon2}. The worldsheet theory obtained in this way admits the interpretation of string theory on a background that interpolates between AdS$_3$ and a linear dilaton background, resulting in a solvable deformation of AdS$_3$/CFT$_2$ that exhibits Hagedorn spectrum at high energy and Cardy spectrum at low energy. This theory has been studied in \cite{Giveon:2017nie, Giveon2, Giribet:2017imm, Giveon3, Babaro:2018cmq, Giveon4, Giveon5, Otro1, Otro2, Otro3, Otro4, Otro5, Otro6} and in references thereof. In particular, correlation functions were studied in \cite{Giribet:2017imm, Giveon3, Babaro:2018cmq}, where it was shown that the insertion of the operator (\ref{Prima}) in the correlation functions produces a logarithmic divergence that leads to the renormalization of the primary operators, which consequently acquire an anomalous dimension. In \cite{Giribet:2017imm}, we compute the 2-point function on the sphere geometry and obtained the anomalous dimension explicitly. This provided us with a direct way of determining the spectrum of the theory. In \cite{Babaro:2018cmq}, we extended the computation of \cite{Giribet:2017imm} to the CFT$_2$ with a conformal boundary: We computed there the expectation value of a bulk primary operator on the disk geometry, and we gave a closed expression for such observable, confirming the result for the anomalous dimension derived in \cite{Giribet:2017imm}. We also computed the bulk-boundary and the boundary-boundary 2-point functions on the disk. Here, we extend these results by explicitly computing the bulk-bulk 2-point function on the disk and the $N$-point function on the sphere, whose forms are also shown to be in perfect agreement with the results of \cite{Giribet:2017imm, Giveon3, Babaro:2018cmq}.

The paper is organized as follows: In section 2, we introduce the marginal deformation of the $SL(2,\mathbb{R})$ WZW model in presence of a conformal boundary. We discuss the form of the relevant 2-point functions on the disk geometry, we revisit the calculation of the boundary-bulk 2-point function using a method different from the ones used in \cite{Babaro:2018cmq}, and we employ the same method to compute the bulk-bulk 2-point function. In section 3, we discuss the theory on the sphere: We compute the 3-point function and the form the $N$-point functions take in the deformed CFT.

\section{The theory on the disk}

\subsection{The action}

The conformal field theory is defined by an action of the form $S=S_{\text{WZW}}+S_{\text{D}}+S_{b}$, consisting of the action of the level-$k$ $SL(2,\mathbb{R})$ WZW theory, the marginal deformation $S_D$, and the appropriate boundary action $S_b$. The WZW action can be written using the Wakimoto fields, namely
\begin{equation}
S_{\text{WZW}}=\frac{1}{2\pi}\int\limits_\Gamma d^2z\, g^{1/2}
\left(\partial\phi\bar\partial\phi+\beta\bar\partial\gamma+\bar\beta\partial\bar\gamma
+\frac{b}{4}R\phi-b^2\, \beta\bar\beta e^{2b\phi}
\right) , \label{TheS}
\end{equation}
with $k=2+b^{-2}$. In terms of this representation, the WZW theory consists of a scalar field with non trivia background charge coupled to a $\beta - \gamma$ ghost system. In this language, the term in the action that realizes the marginal deformation (\ref{Prima}) takes the form
\begin{equation}
S_{{D}}=-\frac{\lambda_0}{\pi}\int\limits_{\Gamma}d^2z\,g^{1/2}\beta\bar\beta, \label{SD}
\end{equation} 
which is controlled by a dimensionless coupling constant $\lambda _0$. $\Gamma$ is the Riemann surface corresponding to the disc geometry, which can be mapped to the complex upper half plane, which can be parameterized by $y\geq 0$ with $z=x+iy$, with its boundary $\partial \Gamma$ being the real line $z=x$. The boundary action $S_b$ is given by
\begin{equation}
S_{{b}}=\frac{1}{4\pi}\int\limits_{\partial\Gamma}dx\ g^{1/4}\left(2b K\phi+{i}
\beta(\gamma+\bar\gamma)-{i\zeta}\beta e^{b\phi}-{i\lambda_b}\beta\right), \label{Slab}
\end{equation}
where $\zeta $, $\lambda _b$ are two arbitrary constants, the latter corresponding to the marginal deformation on the conformal boundary $\partial \Gamma$.

The variation of the boundary terms, using the constraint $\delta(\beta+\bar\beta)|_{z=\bar z}=0$, yields
\begin{equation}
\delta S_{{b}}=\frac{i}{4\pi}\int\limits_{\partial\Gamma}dx
\Big(\delta\phi\left((\bar\partial-\partial)\phi-\zeta b\beta e^{b\phi}\right)+\delta\beta
\left(\gamma+\bar\gamma-\zeta e^{b\phi}-\lambda_b\right)\Big)
\end{equation}
from which we obtain the gluing conditions
\begin{eqnarray}
 \beta+\bar\beta|_{z=\bar z}=0 \ , \ \  \ (\bar\partial-\partial)\phi|_{z=\bar z}=\zeta b\beta e^{b\phi}\ , \ \ \ \gamma+\bar\gamma|_{z=\bar z}=\zeta e^{b\phi}+\lambda_b,  \label{gluing1}
\end{eqnarray}
valid at $\partial \Gamma$, where $z=\bar z$, as the subscript indicates. These gluing conditions are consistent with the symmetry preserving boundary conditions
\begin{eqnarray}
J^{-}+\bar J^{-}|_{z=\bar z}=0\ , \ \ \  T(z)-\overline T(\bar z)|_{z=\bar z}=0 \ . \label{gluing2}
\end{eqnarray}

\subsection{Conformal covariance}

We consider bulk primary vertex operators of the form
\begin{equation}
\Phi^{j}(p|z)=Z_0\,  e^{p\gamma(z)-\bar p\bar\gamma(\bar z)}
e^{2b(j+1)\phi(z,\bar z)}
\end{equation}
where $p $ can be regarded as a complex momentum in the $\gamma $ direction, and $j$ is the momentum in the $\phi $ direction. It is convenient to fix the normalization as $Z_0=|p|^{2(j+1)}$. 

In the undeformed WZW theory these operators have holomorphic and antiholomorphic conformal dimensions $h_j=\bar{h}_j=-b^2 j(j+1)$. Due to the presence of (\ref{Prima}), we expect the conformal dimension to be corrected in the deformed theory 
\begin{equation}
h_j\rightarrow h_{\Phi}^{j,p}=h_j+\delta h_{\Phi}^{p}.
\end{equation}
We also consider the boundary operators
\begin{equation}
\Psi^l( q|\tau)=| q|^{l+1} e^{\tfrac{1}{2} q\gamma(\tau)-\tfrac{1}{2} q\bar\gamma(\tau)}
e^{b(l+1)\phi(\tau)}
\end{equation} 
which are inserted at the boundary with real coordinate $\tau$. In the undeformed theory these operators have conformal dimension $h_l=-b^2 l(l+1)$ and we expect that dimension to be corrected as well, namely 
\begin{equation}
h_l\rightarrow h_{\Psi}^{l, q}=h_l+\delta h_{\Psi}^{ q}.
\end{equation}

There are three correlation functions that have a fully determined dependence on the worldsheet coordinates from conformal invariance (see {\it e.g.} \cite{Fateev:2000ik}). These are the bulk 1-point correlator
\begin{equation}\label{1pointB}
\langle \Phi^{j}(p|z)\rangle\sim \frac{1}{|z-\bar z|^{2h_{\Phi}^{j,p}}}\, ,
\end{equation}
the boundary-boundary 2-point correlator
\begin{equation}\label{2pointbb}
\langle\Psi^l( q|\tau_1)\Psi^l(- q|\tau_2)\rangle\sim
\frac{1}{|\tau_1-\tau_2|^{2 h_{\Psi}^{l, q}}}\, ,
\end{equation}
and the boundary-bulk 2-point correlator
\begin{equation}\label{2pointBb}
\langle \Phi^{j}(p|z) \Psi^l( q|\tau)\rangle\sim
\frac{1}{|z-\bar{z}|^{2h_{\Phi}^{j, p}-h_{\Psi}^{l, q}}|z-\tau|^{2 h_{\Psi}^{l, q}}}\, ;
\end{equation}
and there is a fourth one with a partially determined dependence on the worldsheet coordinates, namely the bulk-bulk 2-point correlator \cite{Herzog:2017xha}
\begin{equation}\label{2pointBB}
\langle \Phi^{j}( p_1|z_1) \Phi^{j}( p_2|z_2)\rangle\sim
\frac{\zeta^{-h_{\Phi}^{j, p_1}-h_{\Phi}^{j, p_2}}}{|z_1-\bar{z}_1|^{2 h_{\Phi}^{j, p_1}} |z_2-\bar{z}_2|^{2 h_{\Phi}^{j, p_2}}} G^{j}_{ p_1 p_2}(\eta);
\end{equation}
see Figure 1. In (\ref{2pointBB}) we defined two projective invariant cross ratios of the four points $z_1$, $\bar{z}_1$, $z_2$ and $\bar{z}_2$ as follows
\begin{equation}
\eta=\frac{(z_1-z_2)(\bar{z}_1-\bar{z}_2)}{(z_1-\bar{z}_2)(\bar{z}_1-z_2)},\qquad
\zeta=\frac{(z_1-z_2)(\bar{z}_1-\bar{z}_2)}{(z_1-\bar{z}_1)(\bar{z}_2-z_2)}
\end{equation}
which are related through $\zeta={\eta}/({1-\eta})$. Notice also the identity $|z_1-\bar{z}_2|^2=|z_1-z_2|^2+|z_1-\bar{z}_1||z_2-\bar{z}_2|$.
\begin{figure}
\includegraphics[width=6.0in]{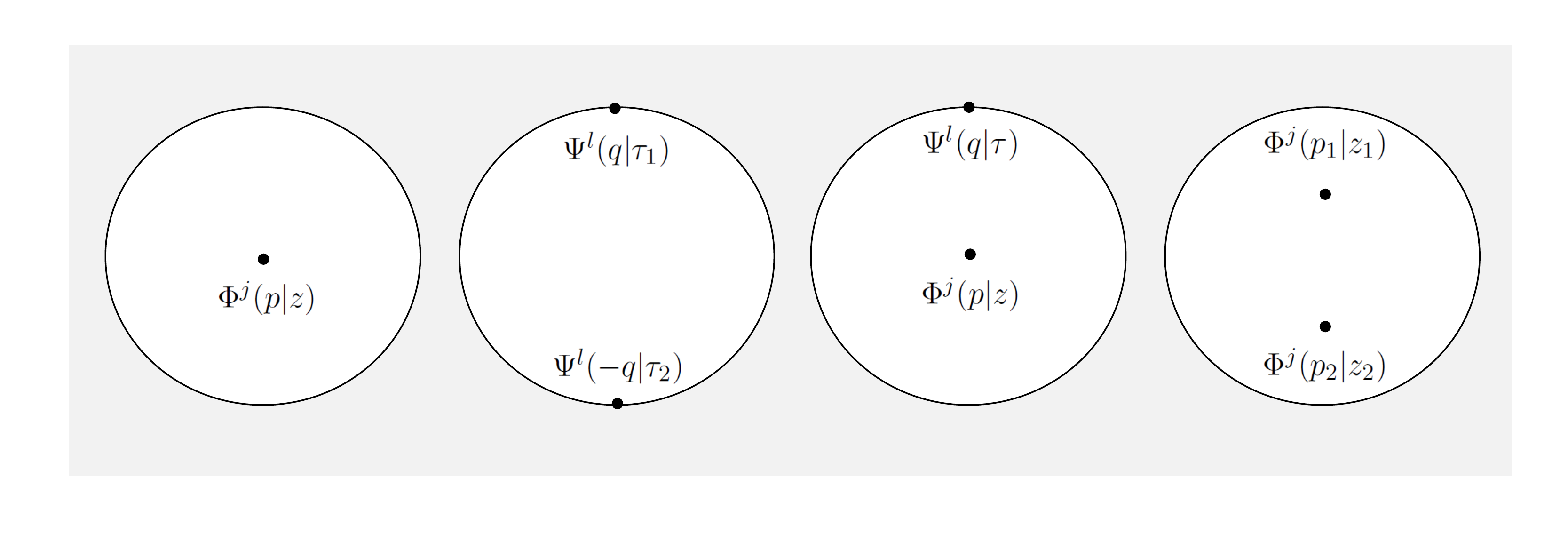}
\caption{Correlation functions on the disk geometry.}
\label{Figure1}
\end{figure}

Our strategy in our previous paper \cite{Babaro:2018cmq} was the following: By carefully treating the deformation (\ref{SD}) in the path integral approach of the 1-point bulk correlator and the boundary-boundary 2-point correlator, we read $\delta h_{\Phi}^{ p}$ and $\delta h_{\Psi}^{ q}$ from the expected scalings (\ref{1pointB})-(\ref{2pointbb}), in a similar way as it was done in \cite{Giribet:2017imm} for the sphere 2-point function. In this way, we obtained the corrections
\begin{equation}\label{anomalous}
\delta h_{\Phi}^{p}=2\lambda_0| p|^2,\qquad 
\delta h_{\Psi}^{q}=2\lambda_0 q^2
\end{equation}
Then, with $\delta h_{\Phi}^{ p}$ and $\delta h_{\Psi}^{ q}$ at hand, we could verify the scaling (\ref{2pointBb}) and verify the consistency of our computation, regularization and renormalization. Now that we have (\ref{anomalous}), we can also verify the scaling (\ref{2pointBB}), which is something we did not address in \cite{Babaro:2018cmq}. Before doing that computation explicitly, we will redo the computation of the bulk-boundary 2-point function (\ref{2pointBb}) in a different way, in order to introduce a new technique that will help us to solve the conformal integrals involved in the bulk-bulk 2-point function.

\subsection{Boundary-bulk two point function revisited}

We will use the path integral techniques of reference \cite{Hikida}. The starting point is to perform the integration over the fields $\gamma$, $\bar\gamma$. This yields two Dirac deltas that constrain the $\beta$, $\bar \beta $ fields. A solution with the proper boundary conditions on the disk only exists for $ p+\bar p+ q=0$ and it is given by
\begin{eqnarray}
\beta(w)&=&\frac{ p}{w-z}+\frac{\bar p}{w-\bar z}+\frac{ q}{w-\tau},\quad \nonumber \\
\bar\beta(\bar w)&=&-\frac{\bar p}{\bar w-\bar z}-\frac{ p}{\bar w-z}-\frac{ q}{\bar w-\tau}.
\end{eqnarray}
Instead of regrouping the denominators, we will multiply term by term in $\beta(w)\bar{\beta}(\bar w)$ producing a total of nine separate contributions to the deformation of the action. Three out of those nine contributions are trivial since they become tadpole-like integrals in dimensional regularization. For example, the first term in $\beta(w)$ times the first term in $\bar\beta(\bar w)$ combines with second respective terms of $\beta$ and $\bar\beta$, and it ultimately contributes with the integral
\begin{equation}
| p|^2\int\limits_{\Gamma} d^{2}w\left(\frac{1}{|w-z|^2}+\frac{1}{|w-\bar{z}|^2}\right)
\end{equation}
where $\Gamma$ is the upper complex half plane. Since the integrand is invariant under $w\to\bar w$ which exchanges the upper half with the lower half, the integral can be carried out in the whole complex plane, giving  
\begin{equation}
\frac{| p|^2}{2}\int\limits_{\mathbb{C}} d^{2}w\left(\frac{1}{|w-z|^2}+\frac{1}{|w-\bar{z}|^2}\right)=
| p|^2\int\limits_{\mathbb{C}} d^{2}w\frac{1}{|w|^2}\to
| p|^2 (l^2 e^{\gamma} \pi)^\epsilon\int d^{2-2\epsilon}w\frac{1}{|w|^2}\stackrel{!}{=}0.
\end{equation}
{where in the last steps we have introduced a regularized version of the integral and set it to zero. As we discuss in the appendix, the resulting integral coming from the products $\int d^2 w\beta(w)\bar{\beta}(\bar{w})$, when considered as a whole, is logarithmically divergent when $w$ hits the insertion points, as one would have expected. Spurious IR and tadpole-like divergences --as the one we have just showed-- will appear in the intermediate steps of the computation as an artifact of the separation of the $\beta(w)\bar{\beta}(\bar{w})$ product in multiple terms. It is a well known fact that using dimensional regularization is a suitable method for handling this kind of divergence mixing.}
The remaining six contributions to the deformation can be reordered in the following way
\begin{equation}\label{defIBb}
-S_{D}=\frac{\lambda_0}{\pi}\int\limits_{\Gamma}d^2 w\beta(w)\bar\beta(w)\rightarrow
-\frac{\lambda_0}{\pi}\left[I_{Bb}^{(1,\epsilon)}(z)
+\left(I_{Bb}^{(1,\epsilon)}(z)\right)^{*}+I_{Bb}^{(2,\epsilon)}(z,\tau)
+\left(I_{Bb}^{(2,\epsilon)}(z,\tau)\right)^{*}\right]
\end{equation}
where we defined the regularized integrals (extension to the whole complex plane using the symmetry $w\to\bar w$ will always be carried out)
\begin{equation}\label{defIBb01}
I_{Bb}^{(1,\epsilon)}(z)=\frac{ p^2}{2}(l^2 e^{\gamma} \pi)^\epsilon
\int d^{2-2\epsilon}w\frac{1}{(w-z)(\bar w-z)}
\end{equation}
\begin{equation}\label{defIBb02}
I_{Bb}^{(2,\epsilon)}(z,\tau)=\frac{ p q}{2}(l^2 e^{\gamma} \pi)^\epsilon
\int d^{2-2\epsilon}w\left(\frac{1}{(w-z)(\bar{w}-\tau)}+\frac{1}{(w-\tau)(\bar{w}-z)}\right).
\end{equation}
Symbol $(\, )^*$ in (\ref{defIBb}) stands for the complex conjugate (recall $ p$ and $z$ are complex while $ q$ and $\tau$ are real). We have introduced a scale $l$ to keep the contribution adimensional and the factor $e^{\gamma\epsilon}\pi^\epsilon$ to absorb irrelevant constants following the same regularization as in our previous paper. {The precise way of regularizing and computing these integrals is explained in the appendix. Here we just quote the results}
%
%We multiply and divide the integrand of (\ref{defIBb01}) by the complex conjugate and we may simplify it, namely
%\begin{equation}
%\frac{ p^2}{2}(l^2 e^{\gamma} \pi)^\epsilon
%\int\limits_{\mathbb{C}} d^{2-2\epsilon}w\frac{(\bar w-\bar z)( w-\bar z)}{|w-z|^2|w-\bar{z}|^2}=
%\frac{ p^2(z-\bar z)}{2}(l^2 e^{\gamma} \pi)^\epsilon
%\int\limits_{\mathbb{C}} d^{2-2\epsilon}w\frac{( w-\bar z)}{|w-z|^2|w-\bar{z}|^2}\nonumber
%\end{equation}
%where we used $(\bar w-\bar z)( w-\bar z)=(z-\bar z)( w-\bar z)+|w-\bar z|^2$ and discarded the second term as it produces a tadpole-like integral. The last remaining integral will be our ``model'' integral, which will appear recurrently along the calculation. We just state the result here and explain its derivation in the appendix:
\begin{equation}\label{defIBb01}
I_{Bb}^{(1,\epsilon)}(z,\tau)=-\frac{ p^2}{2}\pi\,l^{2\epsilon}G_{11}(\epsilon)\frac{1}{|z-\bar{z}|^{2\epsilon}},\qquad
I_{Bb}^{(2,\epsilon)}(z,\tau)=- p q\pi\,l^{2\epsilon}G_{11}(\epsilon)\frac{1}{|z-\tau|^{2\epsilon}} . 
\end{equation}
with $G_{11}(\epsilon)=\tfrac{e^{\gamma\epsilon}\,\Gamma(-\epsilon)^2\Gamma(1+\epsilon)}{\Gamma(-2\epsilon)}$. Combining these two results with their complex conjugates in (\ref{defIBb}), we obtain
\begin{align}
-S_D= & \lambda_0\, G_{11}(\epsilon)\,l^{2\epsilon}\left(
\frac{ p^2+\bar{ p}^2}{2|z-\bar{z}|^{2\epsilon}}+\frac{ p q+\bar{ p} q}{|z-\tau|^{2\epsilon}}
\right)\nonumber\\
= & \lambda_0\left(
\frac{1}{\epsilon}( q^2+2| p|^2)+(2 q^2-4| p|^2)\log\tfrac{|z-\bar{z}|}{l}-4 q^2\log\tfrac{|z-\tau|}{l}+\mathcal{O}(\epsilon)
\right),
\end{align}
where in the last expression we have expanded in $\epsilon$ and made use of $ p+\bar{ p}+ q=0$.
Thus, by exponentiating we find
\begin{equation}\label{BbExponential}
e^{-S_D}=\frac{e^{(  q^2+2| p|^2)\lambda_0/\epsilon}}{|z-\bar{z}|^{4\lambda_0| p|^2-2\lambda_0 q^2}|z-\tau|^{4\lambda_0 q^2}}
\end{equation}
which is exactly both what we expected to obtain ({\it c.f.} (\ref{2pointBb})) to obtain and what we had obtained in the last paper through a different method. This result is consistent with our previous knowledge of the anomalous dimensions $\delta h_{\Phi}^{ p}=2\lambda_0| p|^2$ and $\delta h_{\Psi}^{ q}=2\lambda_0 q^2$. Moreover, the renormalization of the bulk and boundary operators that is consistent with the 1-point bulk and 2-point boundary functions exactly agrees with the one we need now to cancel the poles in (\ref{BbExponential}) and drop the regulator; namely 
\begin{equation}\label{renormalization}
\Phi^{j}( p| z)\rightarrow \Phi^{j}( p| z) e^{-\tfrac{2\lambda_0| p|^2}{\epsilon}}
,\qquad \Psi^l( q|\tau)\rightarrow \Psi^l( q|\tau) e^{-\tfrac{\lambda_0 q^2}{\epsilon}}.
\end{equation}
In all, the correlator computation would lead to the relation
\begin{equation}
\langle \Phi^{j}( p|z) \Psi^l( q|\tau)\rangle_{D}=
\frac{1}
{|z-\bar{z}|^{2\delta h_{\Phi}^{ p}-\delta h_{\Psi}^{ q}}
|z-\tau|^{2 \delta h_{\Psi}^{ q}}}
\langle \Phi^{j}( p|z) \Psi^l( q|\tau)\rangle_{WZW}
\end{equation}
with $\delta h_{\Phi}^{ p}$ and $\delta h_{\Psi}^{ q}$ as before. The subindex $D$ on the left hand side means that the expectation value is computed in the deformed theory ($\lambda_0 \neq 0 $), while the subindex $WZW$ on the right hand side refers to the undeformed WZW theory ($\lambda_0 = 0$); see \cite{Fateev:2007wk, Hosomichi:2006pz}.  

\subsection{Bulk-bulk 2-point function}

Now, let us move to the computation of the bulk-bulk 2-point function. Inserting two fields in the bulk with momenta $ p_1$ and $ p_2$ at the points $z_1$ and $z_2$ leads, after integrating in the $\gamma$, $\bar \gamma $ fields, to a solution for the $\beta$, $\bar \beta $ fields consistent with the boundary conditions, given by
\begin{align}
&\beta(w)=\frac{ p_1}{w-z_1}+\frac{ p_2}{w-z_2}
+\frac{\bar{ p}_1}{w-\bar{z}_1}+\frac{\bar{ p}_2}{w-\bar{z}_2}\nonumber\\
&\bar{\beta}(\bar{w})=-\frac{\bar{ p}_1}{\bar{w}-\bar{z}_1}-\frac{\bar{ p}_2}{\bar{w}-\bar{z}_2}
-\frac{ p_1}{\bar w-z_1}-\frac{ p_2}{\bar w-z_2}
\end{align}
subject to the condition $ p_1+\bar p_1+ p_2+\bar p_2=0$. Now, there are twelve non-trivial contributions which can be regrouped in terms of only three different types of integrals in the following way
\begin{align}\label{defIBB}
-S_{D}= & \frac{\lambda_0}{\pi}\int\limits_{\Gamma}d^2 w\beta(w)\bar\beta(w)\rightarrow
-\frac{\lambda_0}{\pi}\left[I_{BB}^{(1,\epsilon)}(z_1,z_2)
+\left(I_{BB}^{(1,\epsilon)}(z_1,z_2)\right)^{*}+I_{BB}^{(2,\epsilon)}(z_1,z_2)
\right.\nonumber\\
&\left.
+\left(I_{BB}^{(2,\epsilon)}(z_1,z_2)\right)^{*}+I_{BB}^{(3,\epsilon)}(z_1| p_1)+
\left(I_{BB}^{(3,\epsilon)}(z_1| p_1)\right)^{*}+I_{BB}^{(3,\epsilon)}(z_2| p_2)+
\left(I_{BB}^{(3,\epsilon)}(z_2| p_2)\right)^{*}
\right]
\end{align}
where
\begin{equation}\label{defIBB01}
I_{BB}^{(1,\epsilon)}(z_1,z_2)=\frac{ p_1\bar p_2}{2}(l^2 e^{\gamma} \pi)^\epsilon
\int d^{2-2\epsilon}w\left(\frac{1}{(w-z_1)(\bar{w}-\bar{z}_2)}+\frac{1}{(w-\bar{z}_2)(\bar{w}-z_1)}\right)
\end{equation}
\begin{equation}\label{defIBB02}
I_{BB}^{(2,\epsilon)}(z_1,z_2)=\frac{ p_1 p_2}{2}(l^2 e^{\gamma} \pi)^\epsilon
\int d^{2-2\epsilon}w\left(\frac{1}{(w-z_1)(\bar{w}-{z}_2)}+\frac{1}{(w-{z}_2)(\bar{w}-z_1)}\right)
\end{equation}
\begin{equation}\label{defIBB03}
\, \text{and}\, \, I_{BB}^{(3,\epsilon)}(z| p)=\frac{ p^2}{2}(l^2 e^{\gamma} \pi)^\epsilon
\int d^{2-2\epsilon}w\frac{1}{(w-z)(\bar w-z)}.
\end{equation}
These three integrals are of the same kind as those that appeared before and are dealt with in the appendix. Their solution is
\begin{align}
& I_{BB}^{(1,\epsilon)}(z_1,z_2)=-\frac{ p_1\bar p_2\pi\,l^{2\epsilon}G_{11}(\epsilon)}{|z_1-z_2|^{2\epsilon}},\quad I_{BB}^{(2,\epsilon)}(z_1,z_2)=-\frac{ p_1 p_2\pi\,l^{2\epsilon}G_{11}(\epsilon)}{|z_1-\bar{z}_2|^{2\epsilon}} \nonumber\\
\mbox{and}\quad & I_{BB}^{(3,\epsilon)}(z| p)=-\frac{ p^2\pi\,l^{2\epsilon}G_{11}(\epsilon)}{2|z-\bar z|^{2\epsilon}}
\end{align}
Inserting these results in (\ref{defIBB}) and expanding in powers of $\epsilon$, we obtain for the pole and finite piece; namely
\begin{equation}
-{S_D}|_{\epsilon^{-1}}=\frac{\lambda_0}{\epsilon}\left(2| p_1|^2+2| p_2|^2\right)\label{GG1}
\end{equation}
and
\begin{align}
-{S_D}|_{\epsilon^{0}}=& \lambda_0\left[
-2\left(| p_1|^2+| p_2|^2\right)\log\zeta -4| p_1|^2\log\tfrac{|z_1-\bar{z}_1|}{l}
-4| p_2|^2\log\tfrac{|z_2-\bar{z}_2|}{l}
\right.\nonumber\\
& \left.
-\left(( p_1+ p_2)^2+(\bar p_1+\bar p_2)^2\right)\log\eta+\left( p_1^2+ p_2^2+\bar{ p}_1^2+\bar{ p}_2^2\right)\log(1-\eta)
\right]\label{GG2}
\end{align}
where in both cases we used the condition $ p_1+ p_2+\bar p_1+\bar p_2=0$. The subindices $\epsilon^{0}$ and $\epsilon^{-1}$ in (\ref{GG1}) and (\ref{GG2}) refer to the different orders in the expansion in $\epsilon $. The singular piece of the exponential $e^{-{S_D}|_{\epsilon^{-1}}}$ gets canceled with the same renormalization (\ref{renormalization}). Furthermore, the exponential of the remaining finite piece yields
\begin{equation}
e^{-{S_D}|_{\epsilon^{0}}}=\frac{\zeta^{-\delta h_{\Phi}^{ p_1}-\delta h_{\Phi}^{ p_2}}}{|z_1-\bar{z}_1|^{2 \delta h_{\Phi}^{ p_1}} |z_2-\bar{z}_2|^{2 \delta h_{\Phi}^{ p_2}}} {G_D}_{ p_1 p_2}(\eta)
\end{equation}
with
\begin{equation}
{G_D}_{ p_1 p_2}(\eta)=\frac{(1-\eta)^{\lambda_0( p_1^2+ p_2^2+\bar{ p}_1^2+\bar{ p}_2^2)}}{\eta^{\lambda_0( p_1+ p_2)^2+\lambda_0(\bar p_1+\bar p_2)^2}}
\end{equation}
which is the exact behavior we expected according to (\ref{2pointBB}). This manifestly shows the consistency of the calculation of $\delta h_{\Phi}^{ p}$ proposed in \cite{Giribet:2017imm} with the disk 2-point functions. 

\section{Back to the sphere}

\subsection{Three-point function}

Now that we have succeeded with this method to solve new observables in the disk geometry, it is tempting to go back to the sphere and tackle the 3-point function. Conformal symmetry demands the 3-point function to scale like
\begin{align}
&\langle \Phi^{j_1}( p_1|z_1) \Phi^{j_2}( p_2|z_2) \Phi^{j_3}( p_3|z_3)\rangle\sim
\nonumber\\
& \frac{1}
{|z_{12}|^{2 h_{\Phi}^{j_1, p_1}+2 h_{\Phi}^{j_2, p_2}-2 h_{\Phi}^{j_3, p_3}}
|z_{23}|^{2 h_{\Phi}^{j_2, p_2}+2 h_{\Phi}^{j_3, p_3}-2 h_{\Phi}^{j_1, p_1}}
|z_{13}|^{2 h_{\Phi}^{j_1, p_1}+2 h_{\Phi}^{j_3, p_3}-2 h_{\Phi}^{j_2, p_2}}
}
\end{align}
where $z_{ij}=z_i-z_j$. The solutions for the $\beta$ fields after functional integration over $\gamma$ fields is
\begin{eqnarray}
\beta(w)&=&\frac{ p_1}{w-z_1}+\frac{ p_2}{w-z_2}+\frac{ p_3}{w-z_3}\nonumber\\
\bar\beta(\bar w)&=&-\frac{\bar p_1}{\bar w-\bar z_1}-\frac{\bar p_2}{\bar w-\bar z_2}-\frac{\bar  p_3}{\bar w-\bar z_3}
\end{eqnarray}
with $p_1+p_2+p_3=0$. Similarly to the computations before, when making the product $\beta\bar\beta$ three out of the nine terms become trivial. The six remaining terms contribute to the deformation of the action in the following way
\begin{align}
-S_D=-\frac{\lambda_0}{\pi}&\left[
I^{\epsilon}_{ p_1\bar p_2}(z_1,z_2)+I^{\epsilon}_{ p_1\bar p_3}(z_1,z_3)+I^{\epsilon}_{ p_2\bar p_1}(z_2,z_1)\right.\nonumber\\
+&\left.
I^{\epsilon}_{ p_2\bar p_3}(z_2,z_3)+I^{\epsilon}_{ p_3\bar p_1}(z_3,z_1)+I^{\epsilon}_{ p_3\bar p_2}(z_3,z_2)\right]
\end{align}
with
\begin{equation}\label{defIntegralGeneric}
I^{\epsilon}_{ p_i\bar p_j}(z_i,z_j)=
 p_i\bar p_j (l^2 e^{\gamma} \pi)^\epsilon\int d^{2-2\epsilon}w\frac{1}{(w-z_i)(\bar w-\bar z_j)}=
-\frac{ p_i\bar{ p}_j \pi l^{2\epsilon} G_{11}(\epsilon)}{|z_i-z_j|^{2\epsilon}}.
\end{equation}
This is the same kind of integral we have been doing all along. Combining all the contributions we get
\begin{equation}
-S_D=\lambda_0 G_{11}(\epsilon)l^{2\epsilon}\left(
\frac{ p_1\bar  p_2+ p_2\bar  p_1}{|z_{12}|^{2\epsilon}}+
\frac{ p_2\bar  p_3+ p_3\bar  p_2}{|z_{23}|^{2\epsilon}}+
\frac{ p_1\bar  p_3+ p_3\bar  p_1}{|z_{13}|^{2\epsilon}}
\right).
\end{equation}
Expanding in $\epsilon$ and making use of $ p_1+ p_2+ p_3=0$ we obtain for the pole and leading finite piece
\begin{equation}
{-{S_D}|_{\epsilon^{-1}}}=\frac{2\lambda_0}{\epsilon}(| p_1|^2+| p_2|^2+| p_3|^2),
\end{equation}
and
\begin{align}
{-{S_D}|_{\epsilon^{0}}}=-4\lambda_0 &\left[
(| p_1|^2+| p_2|^2-| p_3|^2)\log\tfrac{|z_{12}|}{l}
+(| p_2|^2+| p_3|^2-| p_1|^2)\log\tfrac{|z_{23}|}{l}
\right.\nonumber\\
&\left.+(| p_1|^2+| p_3|^2-| p_2|^2)\log\tfrac{|z_{13}|}{l}
\right].
\end{align}
The exponential of the poles piece $e^{-{S_D}|_{\epsilon^{-1}}}$ gets canceled by the renormalization of the fields, while the exponential of the finite piece is
\begin{equation}
e^{-{S_D}|_{\epsilon^{-0}}}=
\frac{1}
{|z_{12}|^{2 \delta h_{\Phi}^{ p_1}+2 \delta h_{\Phi}^{ p_2}-2 \delta h_{\Phi}^{ p_3}}
|z_{23}|^{2 \delta h_{\Phi}^{ p_2}+2 \delta h_{\Phi}^{ p_3}-2 \delta h_{\Phi}^{ p_1}}
|z_{13}|^{2 \delta h_{\Phi}^{ p_1}+2 \delta h_{\Phi}^{ p_3}-2 \delta h_{\Phi}^{ p_2}}
}
\end{equation}
which is exactly what we expected to obtain: This confirms that both the anomalous dimension and the vertex operators renormalization derived in \cite{Giribet:2017imm} are consistent with the 3-point function. This raises the question as to whether the same occurs for $N$-point functions with $N>3$. 

\subsection{Four-point function}

Consider now the four-point function
\begin{equation}\label{4pointsRelation}
\langle \prod_{i=1}^4 \Phi^{j_i}( p_i|z_i) \rangle_D=\prod_{i<j}|z_{ij}|^{4\lambda_0(p_i\bar{p}_j+p_j\bar{p}_i)}\, \langle \prod_{r=1}^4 \Phi^{j_r}( p_r|z_r) \rangle_{\text{WZW}}
\end{equation}
Let us define the cross-ratios 
\begin{equation}
x=\frac{z_{12}z_{34}}{z_{13}z_{24}} \, , \ \  \bar x=\frac{\bar z_{12}\bar z_{34}}{\bar z_{13}\bar z_{24}}. 
\end{equation}
Following the notation of \cite{diFrancesco}, from the four-point function we can define three types of functions of $x$ and $\bar{x}$, representing three different channels; namely: the $s$-type
\begin{equation}
G_{34}^{21,D}(x,\bar x)=\lim\limits_{z_1,\bar z_1\to\infty}
|z_1|^{4 h_{\Phi}^{j_1,p_1}}
\langle \Phi^{j_1}( p_1|z_1) \Phi^{j_2}( p_2|1) \Phi^{j_3}( p_1|x)  \Phi^{j_4}( p_4|0)\rangle_{D}\, ,\nonumber
\end{equation}
the $t$-type
\begin{equation}
G_{32}^{41,D}(1-x,1-\bar x)=\lim\limits_{z_1,\bar z_1\to\infty}
|z_1|^{4 h_{\Phi}^{j_1,p_1}}
\langle \Phi^{j_1}( p_1|z_1) \Phi^{j_2}( p_2|0) \Phi^{j_3}( p_1|1-x)  \Phi^{j_4}( p_4|1)\rangle_{D}\, , \nonumber
\end{equation}
and the $u$-type
\begin{equation}
G_{31}^{24,D}(\tfrac{1}{x},\tfrac{1}{\bar x})=\lim\limits_{z_4,\bar z_4\to\infty}
|z_4|^{4 h_{\Phi}^{j_4,p_4}}
\langle \Phi^{j_1}( p_1|0) \Phi^{j_2}( p_2|1) \Phi^{j_3}( p_1|\tfrac{1}{x})  \Phi^{j_4}( p_4|z_4)\rangle_{D}\, . \nonumber
\end{equation}
Notice that in the three cases, the powers of $|z_1|$ and $|z_4|$, which arise to render the limit finite, correspond to the conformal dimension of the deformed theory, $h_{\Phi}^{j,p}=h_j+2\lambda_0 |p|^2$. With these definitions, we can use (\ref{4pointsRelation}) to connect these functions with the three different channels of the WZW 4-point function, and then use the crossing symmetry relations for the latter. The crossing symmetry of the non-compact WZW theory has been proven in \cite{Teschner}. This can be expressed as follows
\begin{align}
& G_{34}^{21,\text{WZW}}(x,\bar x)=G_{32}^{41,\text{WZW}}(1\!-\!x,1\!-\!\bar x)\label{CrossingAssumption1}\\
& G_{34}^{21,\text{WZW}}(x,\bar x)=|x|^{-4 h_{j_3}}\ G_{31}^{24,\text{WZW}}(1/x,1/{\bar x})\label{CrossingAssumption2}
\end{align}
where the functions $G_{34}^{21,\text{WZW}}(x,\bar x)$ are defined as above but for the undeformed theory $\lambda_0=0$. Using (\ref{4pointsRelation}) it is possible to show that the deformed theory also satisfies the crossing symmetry relations: The first relation is the simplest, and reads 
\begin{equation}\label{Crossing1}
G_{32}^{41,D}(1\!-\!x,1\!-\!\bar x) = G_{34}^{21,D}(x,\bar x).
\end{equation}
Proving the second relation is more subtle as it requires to use a chain of equalities. The final result reads
\begin{equation}
G_{34}^{21,D}(x,\bar x)=|x|^{-4h^{j_3,p_3}_{\Phi}}\ G_{31}^{24,D}(1/x,1/{\bar x})\, .
\end{equation}
To prove this identity one uses the trasformation of the WZW 4-point function under $x\to 1/x$ and the momentum conservation $\sum_{i=1}^4 p_i=0$.

\subsection{$N$-point functions}

It is easy to write down a general formula for the contribution from the action deformation to the $N$-point function. The solution for $\beta$ in that case is \cite{Hikida}
\begin{equation}
\beta(w)=\sum\limits_{i=1}^N\frac{ p_i}{w-z_i},\qquad
\bar\beta(\bar w)=-\sum\limits_{i=1}^N\frac{\bar p_i}{\bar w-\bar z_i}.
\end{equation}
with $\sum_i p_i=0$. When making the product $\beta(w)\bar\beta(\bar w)$ we obtain $N(N-1)$ non trivial contributions given by
\begin{equation}
-S_D=-\frac{\lambda_0}{\pi}\sum\limits_{\substack{i,j=0 \\ i\neq j}}^N
I^{\epsilon}_{ p_i\bar p_j}(z_i,z_j)=
\lambda_0 l^{2\epsilon}G_{11}(\epsilon)
\sum\limits_{1\leq i <j\leq N}\frac{( p_i\bar p_j+ p_j\bar p_i)}{|z_{ij}|^{2\epsilon}}
\end{equation}
where integral $I^{\epsilon}_{ p_i\bar p_j}$ was defined in (\ref{defIntegralGeneric}). With the conservation condition $\sum_i p_i=0$, which follows from the integral over the zero-mode of $\gamma $, one can derive the following properties
\begin{align}
& \sum\limits_{1\leq i <j\leq N} ( p_i\bar p_j+ p_j\bar p_i)=-\sum\limits_{i=1}^N | p_i|^2\quad\mbox{and}
\nonumber\\
&  p_i\bar p_j+ p_j\bar p_i=\Big{|}\sum\limits_{ k\neq i,j}  p_k\Big{|}^2
-| p_i|^2-| p_j|^2
\end{align}
which we use after expanding in $\epsilon$ to write the exponential of the action deformation as
\begin{equation}
e^{-S_D}=
{\exp\left(\frac{2\lambda_0}{\epsilon}\sum\limits_{i=1}^N | p_i|^2\right)}\,
{\prod\limits_{1\leq i <j\leq N}|z_{ij}|^{-4\lambda_0\left(
| p_i|^2+| p_j|^2-|\!\!\sum\limits_{ k\neq i,j}  p_k|^2
\right)}}.
\end{equation}
Again, the pole piece exponential is exactly what we expected in order to cancel the prescribed renormalization of the fields, in perfect agreement. {After renormalizing the fields, we arrive to the general proportionality relation between correlation functions
\begin{equation}\label{NPointProportionality}
\langle \prod_{i=1}^N \Phi^{j_i}( p_i|z_i) \rangle_D=\prod_{i<j}|z_{ij}|^{4\lambda_0(p_i\bar{p}_j+p_j\bar{p}_i)}\, \langle \prod_{r=1}^N \Phi^{j_r}( p_r|z_r) \rangle_{\text{WZW}}
\end{equation}
This proportionality relation guarantees that the deformed theory satisfies global conformal invariance for a CFT with primary fields with corrected dimensions $h^{j,p}_{\Phi}=h_j+\delta h^p_{\Phi}$. This can be seen as a consequence of the global conformal invariance of correlators of primary fields in the WZW theory with dimensions $h_j$. The explicit check can be made by considering the following three operators for each theory
\begin{equation}
%\hat\Lambda_s^{\text{WZW}}=\sum_{i=1}^N\left(z_i^{s+1}\frac{\partial}{\partial z_i}+(s+1)z_i^s h_{j_i}\right),\quad
\hat\Lambda_s^{D}=\sum_{i=1}^N\left(z_i^{s+1}\frac{\partial}{\partial z_i}+(s+1)z_i^s h^{j_i,p_i}_{\Phi}\right)
\end{equation}
with $s=-1,0,1$. One can explicitly verify the global conformal Ward identities
\begin{equation}
\hat\Lambda_s^{D}\langle  \Phi^{j_1}( p_1|z_1)\dots \Phi^{j_N}( p_N|z_N) \rangle_D=0
\end{equation}
as a consequence of WZW Ward identities and
%\begin{equation}
%\hat\Lambda_s^{\text{WZW}}\langle  \Phi^{j_1}( p_1|z_1)\dots \Phi^{j_N}( p_N|z_N) \rangle_{\text{WZW}}=0
%\end{equation}
 thanks to the fact that the power of the proportionality factor in (\ref{NPointProportionality}) satisfies
\begin{equation}
\sum_{j\neq i}2\lambda_0(p_i\bar p_j+p_j\bar p_i)=-2\delta h^{p_i}_{\Phi}
\end{equation}

}

\acknowledgments

The work of G.G. is supported in part by CONICET (Argentina) through grant PIP 1109 (2017). The work of M.L. is supported in part by ANPCyT (Argentina) through grant PICT 1633 (2015). 

\appendix
\section{Conformal integrals}\label{Integrals}

\subsection*{On the nature of divergences}

{Expanding of the product $\beta\bar\beta$ term by term, all along our work we have found two types of divergent integrals; namely
\begin{equation}
T_0(z)=\int\limits_{\mathbb{C}}d^2w\frac{1}{|w-z|^2},\qquad
I_0(z_1,z_2)=\int\limits_{\mathbb{C}}d^2w\frac{1}{(w-z_1)(\bar w-\bar z_2)}
\end{equation}
with $z, z_1$ and $z_2$ properly identified depending on each case. The first integral, $T_0(z)$, is both UV and IR divergent: the integral explodes when $w\to z$ and for large $w$. The integral $I_0(z_1,z_2)$, on the other hand, is UV finite and IR divergent: it is safe when $w\to z_1,z_2$ (as long as $z_1\neq z_2$) but divergent for $w$ large. At first glance, this could seem contradictory, since we expected to reproduce an anomalous dimension computation as an effect of the regularization and renormalization of only UV divergences. To understand this apparent contradiction consider the solution for $\beta(w)$ for the $N$ point function 
\begin{equation}
\beta(w)=\sum\limits_{i=1}^N\frac{p_i}{w-z_i}=\frac{Q_{N-2}(w)}{(w-z_1)(w-z_2)\dots (w-z_N)}
\end{equation} 
In the last equality we have taken a common factor of all the denominators. Thanks to the conservation condition $\sum_i p_i=0$ the numerator $Q_{N-2}(w)$ is a polynomial of degree $N-2$ (this can also be seen as a consequence of Riemann-Roch theorem limiting the number of zeros of a 1-form). When we insert $\beta$ in the deformed action, we obtain
\begin{equation}\label{app:full}
-S_D=-\frac{\lambda_0}{\pi}\int\limits_{\mathbb{C}} d^{2}w \frac{|Q_{N-2}(w)|^2}
{|w-z_1|^2 |w-z_2|^2\dots |w-z_N|^2}
\end{equation}
By power counting we see that this integral is finite for large $w$ due to the lowered degree of the $Q_{N-2}(w)$ polynomial. On the other hand, the integral has logarithmic divergences when the integration point $w$ approaches the insertion points $z_i$, as expected. From this analysis we see that the mixed UV/IR  and purely IR nature of divergences of $T_0(z)$ and $I_0(z_1,z_2)$ respectively are an artifact of the fact that we are separating the whole integral in $N^2$ different pieces; $N$ of the type $T_0(z)$ and $N(N-1)$ of the type $I_0(z_1,z_2)$, which, when considered all together as in (\ref{app:full}) have the correct divergent behavior.

It is easy to realize that a regularized version of the integral (\ref{app:full}) is much more difficult to attack than a regularized version of $T_0(z)$ and $I_0(z_1,z_2)$. The only reason why we made the separation of $\beta\bar\beta$ into $N^2$ terms is to make the computation simpler. Nevertheless, one could worry that by separating an UV divergent integral as (\ref{app:full}) in multiple terms which create artificial IR divergences we could be following an inconsistent path.
The nice feature about this is that we are using dimensional regularization which is very well suited to treat UV and IR divergences on equal footing and dealing with an artificial mixing of spurious divergences.
}

\subsection*{Regularization and solution}

{Let us go back to integral $I_0$ defined above ($T_0$ is just a tadpole and we put it to zero in dimensional regularization). All the non trivial integrals that appeared in this note have the form}
\begin{align}\label{app:NonRegDefinition}
I_0(z_1,z_2)=&\int\limits_{\mathbb{C}}d^2w\frac{1}{(w-z_1)(\bar w-\bar z_2)}=
\int\limits_{\mathbb{C}}d^2w\frac{(\bar w-\bar z_1)(w-z_2)}{|w-z_1|^2| w- z_2|^2}=\nonumber\\
&(\bar z_2-\bar z_1)\int\limits_{\mathbb{C}}d^2w\frac{w-z_2}{|w-z_1|^2| w-z_2|^2}
=(\bar z_2-\bar z_1)\int\limits_{\mathbb{C}}d^2w\frac{w}{|w|^2| w+z_2-z_1|^2}
\end{align}
with $z_1$ and $z_2$ being two arbitrary complex variables to be properly identified depending on which integral one is facing at. In the last two equalities of this equation we have taken two steps: firstly, we have written $(\bar w-\bar z_1)(w-z_2)=|w-z_2|^2+(\bar z_2-\bar z_1)(w-z_2)$ and we discarded the first piece because it will eventually lead to a tadpole-like integral when properly regularized; secondly, we have shifted the integration $w\to w+ z_2$. Let us focus in the last piece 
\begin{equation}\label{app:NonRegDefinition2}
J_0(z_1,z_2)=\int\limits_{\mathbb{C}}d^2w\frac{w}{|w|^2| w+z_2-z_1|^2}
\end{equation}
and define the following regularized Feynman vector-like bubble integral in $D=2-2\epsilon$ dimensions
\begin{equation}\label{app:RegDefinition3}
J^{\alpha}(x_0)=2(l^2 e^{\gamma} \pi)^\epsilon\int d^{2-2\epsilon}x\frac{x^\alpha}{x^2(x+x_0)^2}
\end{equation}
In $D$ dimensions the vectors $x$ and $x_0$ have $D$ components that reduce to two components when we take $\epsilon\to 0$. We associate those two components with the real and imaginary part of the points $w$ and $z_2-z_1$ in (\ref{app:NonRegDefinition2}). Moreover, the integral $J^{\alpha}(x_0)$ has also $D$ components that reduce to two which we associate with the real and imaginary part of $J_0(z_1,z_2)$.

Integral (\ref{app:RegDefinition3}) is a standard dimensionally regularized Feynman integral which can be easily solved, giving
\begin{equation}
J^{\alpha}(x_0)=-\frac{\pi\,l^{2\epsilon}\,G_{11}(\epsilon)}{(x_0^2)^{1+\epsilon}}\,x_0^{\alpha}
\end{equation}
and by the association explained in the last paragraph this means that the regularized version of $J_0(z_1,z_2)$ becomes
\begin{equation}
J^{(\epsilon)}_0(z_1,z_2)=-\pi\,l^{2\epsilon}\,G_{11}(\epsilon)\frac{(z_2-z_1)}{|z_2-z_1|^{2+2\epsilon}};
\end{equation}
going back to the original integral, its regularized version becomes
\begin{equation}
I^{(\epsilon)}_0(z_1,z_2)=-\frac{\pi\,l^{2\epsilon}\,G_{11}(\epsilon)}{|z_2-z_1|^{2\epsilon}}=
2\pi\left(\frac{1}{\epsilon}-2\log\tfrac{|z_2-z_1|}{l}+\mathcal{O}(\epsilon)\right).
\end{equation}

%\paragraph{Note added.} ...

% The bibliography will probably be heavily edited during typesetting.
% We'll parse it and, using the arxiv number or the journal data, will
% query inspire, trying to verify the data (this will probalby spot
% eventual typos) and retrive the document DOI and eventual errata.
% We however suggest to always provide author, title and journal data:
% in short all the informations that clearly identify a document.

\end{document}